\documentstyle[epsfig]{ichep}
\def\lapproxeq{\lower .7ex\hbox{$\;\stackrel{\textstyle <}{\sim}\;$}}
\def\gapproxeq{\lower .7ex\hbox{$\;\stackrel{\textstyle >}{\sim}\;$}}

\def\GeV{{\rm GeV}}
\def\MeV{{\rm MeV}}

\begin{document}
\title{
\begin{flushright}
DTP/94/82 \\
~~~~ \\
\end{flushright}
Spin-dependent Parton Distributions$^*$}
\author{T.\ Gehrmann$^{\dag}$ and W.J.\ Stirling$^{\dag\S}$}
\address{$^{\dag}$Department of Physics, University of Durham, Durham DH1~3LE, 
England. \\ 
$^{\S}$Department of Mathematical Sciences, 
University of Durham, Durham DH1~3LE, England.}   

\abstract{We perform a global leading-order QCD fit to recent polarized
structure function data in order
to extract a consistent set of spin-dependent parton distributions. 
 Assuming that there is no significant intrinsic
polarization of the quark sea,  the data are  consistent with a
 modest amount of the proton's spin carried by the gluon, although the shape
 of the gluon distribution is not well constrained.
 We show how  inelastic $J/\psi$ production in polarized photon-hadron
 scattering can, in principle, provide definitive 
 information on the shape of the gluon distribution.} 

\twocolumn[\maketitle] 
\fnm{7}{Presented by WJS at the 27th International Conference
on High Energy Physics, Glasgow, July 1994}

Several experiments \cite{slac93,smc93,smc94b} have recently presented new measurements
of the polarized deep-inelastic structure function $g_1$. Combined
with earlier measurements \cite{bau83,emc89}, the data cover a broad range in $x$ and $Q^2$ and provide, 
for the first time, detailed information on the spin-dependent parton
distributions. We present here a summary of a recent analysis \cite{gs94} 
in which
 we perform a leading-order QCD fit to 
the high-$Q^2$ data and extract a consistent set of parton distributions.

\begin{figure}
\begin{center}
~ \epsfig{file=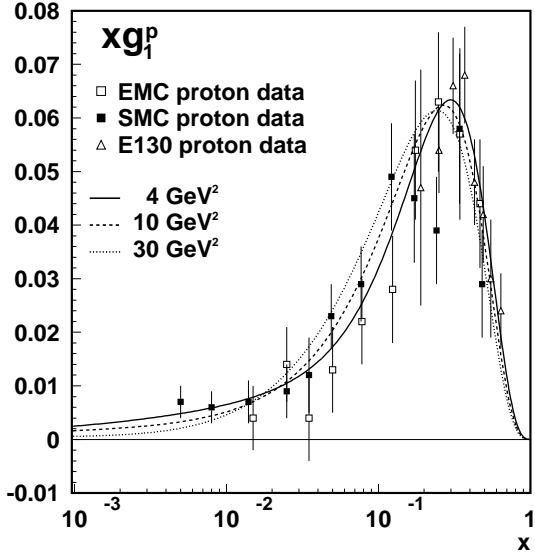,width=7cm}
\caption{Fit to the $g_1^p$ structure function with set A gluon}
\label{fig:proton}
\end{center}
\end{figure}

\begin{figure}
\begin{center}
~ \epsfig{file=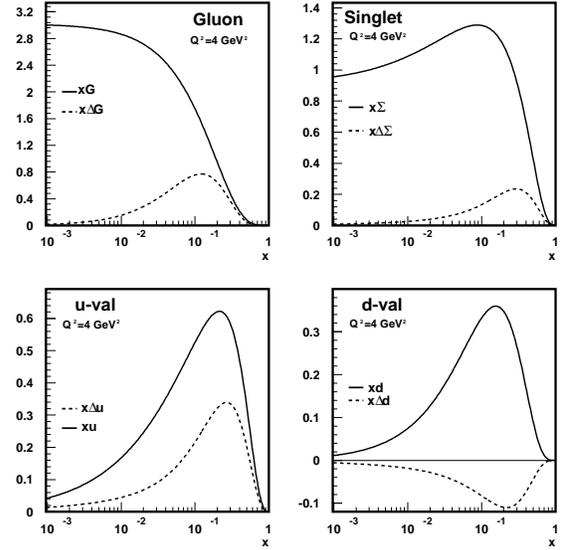,width=8cm}
\caption{The polarized  gluon (set A), quark singlet,
$u_v$ and $d_v$  distributions at $Q_0^2 = 4\ \GeV^2$ obtained
from the fit to the deep inelastic scattering data.}
\label{fig:polpart}
\end{center}
\end{figure}

The fact that the measured value \cite{smc94b} of the integral of $g_1$,
$\Gamma_1^{p} =  0.142 \pm 0.008 \pm 0.011$,
is  less than the Ellis-Jaffe prediction (0.18) \cite{ell74a}
 suggests
that the  gluon makes a significant contribution.
At leading order, we can write \cite{alt88} (for 3 quark flavours)
\[
g_1(x,Q^2) = {1\over 2}\; \sum_{q,\bar q}\; e_q^2
\Delta q(x Q^2) 
- {\alpha_s(Q^2) \over 6\pi} \Delta G(x,Q^2) .
\]
In our model, we assume  that there is no  polarized sea-quark distribution
at $Q^2 = 4\; \GeV^2$.
The only {\it a priori} constraints on the distributions are (i) the
specification of the first moments by the sum-rule and hyperon decay data,
 and (ii)
the requirement of  positivity of the individual helicity distributions, 
$\vert \Delta f \vert \leq f\; , \ (f=q,G)$. In addition, Regge and coherence
arguments can be used to fix the $x\to 0$ behaviour. Unfortunately
neither the neutron \cite{slac93} nor
deuteron \cite{smc93}  $g_1$ data are precise enough to
constrain the   $\Delta d$ distribution: the fit is dominated by $g_1^p$
which depends mainly on  $\Delta u$. 
For consistency, we choose the same $\Lambda_{\rm LO}^{(4)}
= 177 \ \MeV$ and $Q_0 = 2$~GeV values as
\cite{owe91}, and similar starting parametrizations at $Q^2= Q_0^2 $:
\begin{eqnarray*}
x\Delta u_{v} & = & \eta_u A_u x^{a_u}(1-x)^{b_u}(1+\gamma_u x) \nonumber\\
x\Delta d_{v} & = & \eta_d A_d x^{a_d}(1-x)^{b_d}(1+\gamma_d x) \nonumber\\
x\Delta \bar q  &=& 0 \qquad (q=u,d,s,c)     \nonumber \\
x\Delta G & = & \eta_G A_G x^{a_G}(1-x)^{b_G}(1+\gamma_G x)  \nonumber
\label{eq:form}
\end{eqnarray*}
with normalization factors  $A_f$  ($f=q,G$) to ensure that
 $\int_0^1 dx \; \Delta f(x,Q_0^2)  = \eta_f$:
We impose a cut  $Q^2 > 4\; \GeV^2$ on the data to suppress higher-twist
contributions. The fit to the  $g_1^p$ structure function data from
SLAC \cite{bau83}, EMC \cite{emc89} and SMC \cite{smc94b}
is shown in Fig.~\ref{fig:proton} and the resulting
parameters are listed in  Table~\ref{tab:parameters}.
The shape
 of the gluon distribution is not well constrained.
\begin{table}
\begin{center}
\begin{tabular}{|c|r|c|} \hline
Parameter & Value &  Comments \\ \hline
$\eta_u$ & 0.848 & $\eta_u = 2{F}$ \\ 
$b_u$ & 3.64 & $=$ unpolarized $b_u$  \\ 
$a_u$ & 0.46  & fitted  \\ 
$\gamma_u$ & 18.36  & fitted  \\ \hline
$\eta_d$ & $-$0.294 & $\eta_d = {F} - {D}$ \\ 
$b_d$ & 4.64 & $=$ unpolarized $b_d$ \\ 
$a_d$ & 0.46 & $= a_u$ (Regge) \\ 
$\gamma_d$ & 18.36  & $=\gamma_u$ (constrained)  \\ \hline
$\eta_G$ & 1.971 & from $\Gamma_1^p$ \\ 
$b_G$ & 7.44 & fitted \\ 
$a_G$ & 1.0 & coherence arguments \\ 
$\gamma_G$ & 0.0 & constrained \\ \hline
\end{tabular}
\caption{Parameters for the set A partons}
\label{tab:parameters}
\end{center}
\end{table}
In the fit shown in Fig.~\ref{fig:proton}, $\gamma_G = 0$ (the set A gluon)
has been chosen. Equally good fits can be obtained with other values.
To span the range allowed by positivity, we define two other sets 
which have   $\gamma_G = 18.0$ (set B) and  $\gamma_G = -3.5$ (set C), 
with the other parameters fitted to the data.
Fig.~\ref{fig:polpart} shows the set A
gluon, quark-singlet, $u$-valence and $d$-valence
distributions at $Q_0^2$ obtained from the fit. For comparison, 
the unpolarized distributions of Ref.~\cite{owe91} are also shown.
Note that  perturbative
 evolution generates  a polarized sea distribution
for $Q^2 > Q_0^2$. However in the range of $Q^2$ relevant
to the structure function measurements the polarized sea is small in comparison
to $\Delta u$ and $\Delta d$. This appears to be consistent with 
the preliminary measurements of $\Delta\bar q$ from inclusive hadron production,
reported by SMC at this conference \cite{nassalski}.

\begin{figure}
\begin{center}
~ \epsfig{file=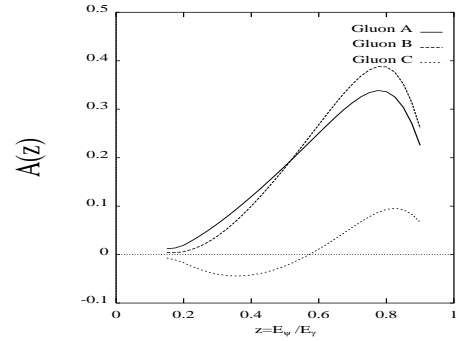,width=6cm}
\caption{The asymmetry ${\cal A}(z)$ predicted by 
 the different gluon distributions, for a 
 photon beam of energy $E_{\gamma}=45$ GeV  on a stationary proton
 target \protect{\cite{bre94}}.}
\label{fig:z}
\end{center}
\end{figure}

Inelastic $J/\psi$ production in polarized photon-nucleon scattering 
provides a possible method of measuring $\Delta G$ \cite{gui88}.
The cross section $d\Delta\sigma^{\gamma N}/{dp_T^2 dz}$
(where $z=E_{J/\psi}/E_{\gamma}$) is proportional to $\Delta G$ in leading
order. The asymmetry ${\cal A}(z)$, defined by integrating over 
$p_T^2 > 0.25$ GeV$^2$, is shown in Fig.~\ref{fig:z} for the three
gluons.


\Bibliography{99}

\bibitem{slac93}
SLAC-E142 collaboration: D.L.~Anthony {\it et al.}, 
Phys. Rev. Lett. {\bf 71} (1993) 959.

\bibitem{smc93}
SMC: B.~Adeva {\it et al.}, Phys. Lett. {\bf B302} (1993) 553.

\bibitem{smc94b}
SMC: D.~Adams {\it et al.}, Phys. Lett. {\bf B329} (1994) 399.

\bibitem{bau83}
SLAC-Yale collaboration: M.J.~Alguard {\it et al.},
Phys. Rev. Lett. {\bf 37} (1976) 1261; G.~Baum {\it et al.},
Phys. Rev. Lett. {\bf 45}  (1980) 2000;  {\bf 51}  (1983) 1135.

\bibitem{emc89}
EMC: J.~Ashman {\it et al.}, Nucl. Phys. {\bf B328} (1989) 1.

\bibitem{gs94} T.~Gehrmann and W.J.~Stirling,
 Durham preprint DTP/94/38 (1994), 
to be published in Zeit. Phys. {\bf C}.

\bibitem{ell74a}
J.~Ellis and R.L.~Jaffe, Phys. Rev. {\bf D9} (1974) 1444, erratum {\bf
D10} (1974) 1669.


\bibitem{alt88}
G.~Altarelli and G.G.~Ross, Phys. Lett. {\bf B212} (1988)  391.

\bibitem{owe91}
J.F.~Owens, Phys. Lett. {\bf B266}  (1991) 126.

\bibitem{nassalski}
SMC: J.~Nassalski, these proceedings.

\bibitem{gui88}
J.Ph.~Guillet, Z. Phys. {\bf C39} (1988) 75.

\bibitem{bre94}
V.~Breton, {\it  Measurement of $\Delta G$ by $J/\psi$ 
Photoproduction at SLAC},
SLAC proposal, 1994.

\end{thebibliography}
\end{document}